\documentclass[conference]{IEEEtran}
\usepackage{cite}
\usepackage{amsmath,amssymb,amsfonts}
\usepackage{algorithmic}
\usepackage{graphicx}
\usepackage{textcomp}
\usepackage[hidelinks]{hyperref}
\usepackage{xcolor}

\begin{document}

\title{2.5D Root of Trust: Securing the Chiplet Ecosystem}

\author{\IEEEauthorblockN{Charles Williams}
\IEEEauthorblockA{Dept. of Electrical and Computer Engineering\\
Texas A\&M University\\
charlesw2000@tamu.edu}
\and
\IEEEauthorblockN{Mohammed Nabeel}
\IEEEauthorblockA{Tandon School of Engineering\\
New York University\\
mohammed.nabeel@nyu.edu}
\and
\IEEEauthorblockN{Gino Chacon}
\IEEEauthorblockA{AheadComputing\\
gino.chacon@aheadcomputing.com}
\and
\IEEEauthorblockN{Ozgur Sinanoglu}
\IEEEauthorblockA{Center for Cyber Security\\
New York University Abu Dhabi\\
ozgursin@nyu.edu}
\and
\IEEEauthorblockN{Paul V. Gratz}
\IEEEauthorblockA{Dept. of Electrical and Computer Engineering\\
Texas A\&M University\\
pgratz@tamu.edu}
\and
\IEEEauthorblockN{Johann Knechtel}
\IEEEauthorblockA{Center for Cyber Security\\ 
New York University Abu Dhabi\\
johann@nyu.edu}
}

\maketitle

\begin{abstract}
  The semiconductor industry is rapidly transitioning from monolithic
  systems-on-chip toward heterogeneous, multi-vendor 2.5D
  chiplet ecosystems integrated via silicon interposers. While this
  paradigm shift offers immense benefits in yield, cost, and
  time-to-market, it radically expands the attack surface. Integrating
  chiplets from untrusted foundries and design houses introduces
  vulnerabilities to hardware Trojans, IP piracy, and system-level
  communication exploits. Critically, chip-level security features and
  conventional Root of Trust (RoT) proposals are insufficient in this
  context: any component, including the interconnect fabric itself, may be
  sourced from an untrusted vendor.
  This perspective paper surveys
  state-of-the-art security strategies for interposer-based 2.5D
  integration, focusing on three threat categories: interconnect
  attacks (snooping, spoofing, and man-in-the-middle), cache coherence
  exploits including complex forging attacks, and microarchitectural
  side-channel threats. We examine design-time defenses via 2.5D split
  manufacturing and, more crucially, runtime defenses that establish
  an active interposer as a physically isolated 2.5D RoT.
  By embedding so-called transaction monitors and coherence
  message checkers within the trusted interposer fabric, the
  system enforces memory access permissions by construction and neutralizes
  coherence-level attacks without need for modifying/securing the commodity
  chiplets. Finally, we review the EDA flows required to realize these
  defenses and show they concurrently improve power and signal
  integrity while reducing overall system footprint.
\end{abstract}

\section{Introduction}

To overcome the CMOS scalability bottleneck, industry is
transitioning from monolithic Systems-on-Chips (SoCs) toward
high-density 2.5D chiplet-based designs. 2.5D integration enables a
modular ``plug-and-play'' ecosystem in which third-party chiplets are
designed and manufactured separately on various process nodes and
integrated onto a silicon
interposer~\cite{loh2021understanding,kim2019architecture}. Multiple products from AMD, Apple, Intel, and Nvidia have adopted chiplet architectures~\cite{naffziger20202,naffziger2021pioneering, loh2021understanding, apple2022m1ultra,smith2025amd,nassif2022sapphire, 9567038, greenhill20173}.

At present, nearly all chiplets in existing 2.5D systems are
designed by one vendor and manufactured by a second. That said, in the
near-term future, the immense economic benefits
means that the design and fabrication of these
chiplets will be outsourced to large degrees to multiple different
companies, some of which may be untrusted~\cite{nabeel2020}.
Consequently, a system designer integrating
commodity chiplets from various vendors will no longer be able to
guarantee the trustworthiness of individual
components~\cite{chacon2024}.  Notably, most existing hardware security features
used in traditional SoCs, such as trusted execution environments (e.g., ARM TrustZone
or Intel SGX), are commonly embedded within untrusted silicon and/or rely
on security mechanisms that do not extend past the boundaries of the
CPU in which they are
implemented~\cite{arm2009security,mckeen2016intel}. This makes them
prime targets for circumvention or malicious modification by supply
chain adversaries~\cite{nabeel2020}.


Most Root of Trust (RoT) proposals for 2.5D integration implement the RoT as a
chiplet that acts as centralized security authority,
managing the TEEs, ensuring authenticity of all chiplets in the
system, and providing identity-centric communication links between
chiplets~\cite{liu2023securing,rambus2026monolithic}. However, if the
chiplet manufacturing process is untrusted, such supposedly secure chiplet 
may itself be compromised. Additionally, the RoT chiplet's
view of the system's operations is limited to the information it
receives from the network-on-chip (NoC), meaning that a malicious chiplet incorporated into
the system can attempt to bypass the RoT's expected system behavior
beyond its observations capabilities. Furthermore, a sophisticated attacker may
introduce a hardware Trojan in the NoC itself to bypass a particular
chiplets' defenses, as the NoC itself may be an untrusted
component~\cite{kulkarni2021packets,charles2021nocattksurvey}.

Overall, securing modern 2.5D systems requires a fundamental shift from
chip-level security to robust, system-level architectures built on a
physically isolated RoT whose trustworthiness is enforced through a
secure and trusted supply chain~\cite{chacon2024}. This survey 
reviews the state-of-the-art in 2.5D chiplet security, focusing
on interposer-based defenses.\footnote{We exclude security considerations for native 3D IC
technologies (e.g., through-silicon via (TSV) stacked logic,
face-to-face bonding, and monolithic 3D) to concentrate on
the unique threat models and architectural opportunities presented by
2.5D integration. For the former, we refer the reader to~\cite{knechtel2019, Knechtel2017, 3062293}.}

\begin{figure}[tb]
    \centering
    \includegraphics[trim={0 15mm 0 60mm},clip,width=\columnwidth]{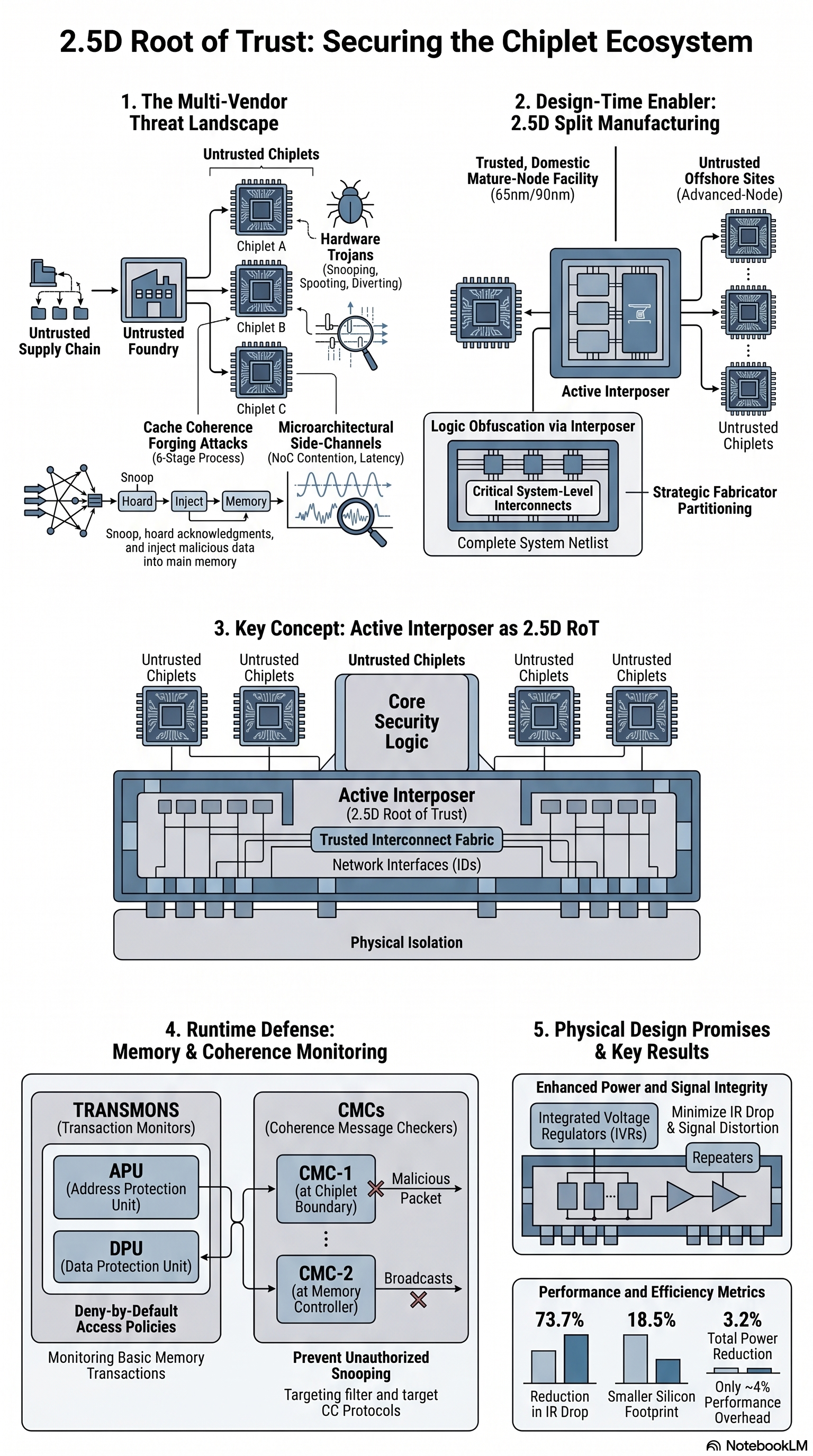}
    \vspace{-2mm}
    \vspace{-2mm}
    \vspace{-2mm}
    \caption{Overview of key concepts and findings of our prior work reviewed in this perspective paper. Image generated using NotebookLM.}
    \vspace{-2mm}
    \label{fig:overview}
\end{figure}

This paper is organized as follows.
In Section~\ref{sec:threats}, we outline prominent threats for chiplet systems.
In Section~\ref{sec:SM}, we briefly introduce split manufacturing for 2.5D integration, providing a defense foundation by construction and also enabling subsequent runtime defenses.
In Section~\ref{sec:TRANSMON}, we review the concept of using an active interposer for a modern 2.5D RoT. We also review its use case for monitoring transactions-based memory access.
In Section~\ref{sec:CC}, we extend the concept of active 2.5D RoT for securing modern cache coherence systems.
In Section~\ref{sec:EDA}, we discuss how to realize an active 2.5D RoT by means of advanced physical design.
We conclude in Section~\ref{sec:conc}.
Figure~\ref{fig:overview} provides an overview of key concepts and findings covered in this review paper.

\section{Threat Landscape in Chiplet Ecosystems}
\label{sec:threats}

Adversaries in multi-chiplet systems can utilize 
malicious software, design bugs, or intentionally inserted Hardware
Trojans~\cite{bhunia2018hardware}.
In order for the IP components within each chiplet to access the
shared memory and other system-level resources, individual chiplets
must share a common interconnect fabric and memory space, making
system-level communication a highly vulnerable attack
vector~\cite{chacon2022,chacon2024,shafkat2022secure,suzano2024hardware}. We
address a threat model in which an attacker seeks to insert a
malicious modification to a chiplet that is capable of exploiting
system-level memory communication behavior.

\subsection{Threats for Interconnects and Shared Memory}

A wide range of works have explored potential attacks occurring
directly on traditional NoC interconnects~\cite{charles2021nocattksurvey,
  basak2017taxonomy,ancajas2014fortnocs,fiorin2008secure}.  Unlike
such traditional threats, however, 2.5D system interconnects
are uniquely vulnerable due to the inclusion of IP components from
multiple design and fab vendors into one package.  In this context,
the interconnect between chiplets becomes a unique attack vector.
In our prior work we show that malicious chiplets can
execute several critical attacks on the system-level NoC that can
impact processing on other chiplets throughout the
system: snooping, spoofing, modifying, diverting, and man-in-the-middle~\cite{nabeel19, nabeel2020, chacon2024}.

\subsection{Threats for Cache Coherence}

Cache coherence (CC) protocols are system-spanning mechanisms that allow
multi-core chiplets to maintain a consistent
view of memory. Coherence mechanisms operate transparently to the
operating system and do not directly enforce virtual or physical
memory permissions, leaving these vulnerable to exploitation by
attackers. For example, a Trojan embedded in a chiplet’s CC controller can
generate legal coherence messages to bypass page-table-based
protections, observing and directly manipulating memory regions that
the compromised chiplet lacks privileges to access~\cite{chacon2024}.

CC protocols like MOESI
Hammer~\cite{conway2010moesihammer} rely on broadcasting
messages to all cores within a system, enabling a malicious chiplet to
spy on the system's state~\cite{chacon2022}. This 
enables complex \textit{Forging Attacks} by a Trojan embedded in the CC
controller of the chiplet, executed in multiple stages~\cite{chacon2022,chacon2024}.
In short, the Trojan snoops traffic to target an address, requests exclusive ownership to trigger global directory invalidations, hoards incoming verification acknowledgments, and injects a rogue writeback to overwrite main memory.
Remarkably, this attack occurs without the Trojanized chiplet ever
owning the victim's memory address space. The 
chiplet will receive no indication that a series of transactions were
triggered on its behalf, and the global CC mechanism operates
as intended, without any indication of the compromise.

Beyond direct memory manipulation, shared CC fabrics expose
chiplets to microarchitectural side-channel threats. A
malicious chiplet can monitor contention and latency variations on the
shared NoC to infer the memory access patterns of co-resident
chiplets, constructing a covert channel without issuing any explicit (unauthorized) request~\cite{chacon2022}.
Microarchitectural speculation
mechanisms present an additional attack surface: a Trojan may
issue speculative or out-of-bounds prefetch requests to addresses
outside its permitted region, leveraging the CC protocol to
trigger state changes or data movements that reveal information about
victim chiplets' working sets~\cite{chacon2024}.

A natural question is whether established NoC security
techniques~\cite{fiorin2008secure,ancajas2014fortnocs,charles2021nocattksurvey}
can be applied to chiplet-based systems. Unfortunately, the
multi-vendor nature of chiplet systems fundamentally undermines
this approach. Traditional techniques assume a
single trusted entity who implements both the interconnect and the
security features embedded within it. In a chiplet system assembled
from outsourced components,
no such assumption holds,
motivating a physically isolated RoT
outside the untrusted interconnect.

\section{Design-Time Defenses: 2.5D Split Manufacturing}
\label{sec:SM}

To mitigate IP piracy and targeted insertion of Trojans,
split manufacturing is highly effective. While 
heavily discussed in the context of 2D ICs and native 3D
ICs~\cite{9344587, 8203796, 3195970, 3317780, cryptography6020022, unsplit, 8587715, patnaik2019modern, imeson2013securing}, it is also valid for 
2.5D integration~\cite{xie2017security,knechtel2019}.

In a 2.5D split manufacturing approach, the designer can
obscure the overall logic by hiding critical
system-level interconnects within the interposer. Prior work 
has demonstrated that passive interposers can conceal
wires from untrusted foundries fabricating the chiplets~\cite{xie2017security}.
Since the front-end-of-line
facilities manufacturing the individual chiplets never see the
interposer's routing, they cannot infer the complete system
netlist, preventing them from understanding the context of the IP
blocks or identifying optimal targets for Trojans~\cite{knechtel2019,patnaik2019modern,8587715,imeson2013securing}.
Furthermore, migrating to an \textit{active}
interposer allows designers to split actual logic across trusted and
untrusted facilities, elevating the security guarantees significantly~\cite{knechtel2019}, as also discussed in detail in the following sections.

\section{Runtime Defenses: Active Interposer as 2.5D RoT and Memory Transaction Monitoring}
\label{sec:TRANSMON}

While passive interposers provide only wiring, \textit{active
  interposers} incorporate logic elements (e.g., NoC routers,
repeaters, integrated voltage regulators), enabling them to function
as a 2.5D RoT~\cite{nabeel19, nabeel2020,
  park2020,chacon2022,chacon2024}. Given that active interposers are
typically manufactured using mature, cost-effective nodes (like 65nm
or 90nm), it is highly realistic to commission a fully trusted,
domestic facility for their fabrication~\cite{park2020,
  nabeel19,nabeel2020,chacon2022,chacon2024}.

The cornerstone of a 2.5D RoT is strict physical separation. Untrusted
commodity chiplets are isolated from the security features; they must
rely entirely on the interposer's interconnect fabric for system-level
communication~\cite{nabeel19,nabeel2020,chacon2022,chacon2024}.
Since the
network interfaces are physically located within the trusted
interposer, untrusted chiplets cannot tamper with identifiers,
rendering spoofing attacks physically impossible.

\subsection{Transaction Monitors}
\label{sec:APU}

In one approach to monitor runtime behavior, the active interposer
embeds so-called Transaction Monitors (TRANSMONs) between chiplets
and the interposer NoC~\cite{nabeel19, nabeel2020}.
A TRANSMON inspects
transactions via two hardware units:
\begin{itemize}
\item \textbf{Address Protection Unit (APU):} Evaluates read/write
  requests against Policy Register Spaces (PRSs). An APU policy
  dictates the Master ID,
  base address,
  address mask,
  and access permissions.
  Transactions targeting unauthorized regions are
  blocked by a Slave Access Filter (SAF), which safely drops the
  transaction and returns an error response to the requester.
\item \textbf{Data Protection Unit (DPU):} Provides fine-grained,
  data-level protection. A DPU policy defines a restricted
  data value
and a data mask
  alongside address bounds. It prevents the leakage
  of sensitive soft assets (e.g., cryptographic keys) into shared
  memory by blocking ``shadow writes,'' and prevents the malicious
  overwriting of shared system semaphores.
\end{itemize}

By enforcing a deny-by-default policy, TRANSMONS prevent any malicious
access. To manage these policies dynamically, a secure 
core (embedded within the active interposer) acts as scheduler and policy compiler, ensuring untrusted
software cannot tamper with the PRSs~\cite{nabeel19, nabeel2020}.

\subsection{Memory-Security Features}

To protect against faults or malicious modifications within shared
memories themselves, TRANSMONs can enforce cryptographic protections
such as Error Correction Codes (ECC) or Cyclic Redundancy Checks
(CRC). Because address handling is governed by the NoC protocol, ECC
results can be fetched in parallel and stored in a separate (trusted)
memory chiplet, preventing overhead latencies during read-out~\cite{nabeel19, nabeel2020}.

\section{Runtime Defenses: Cache Coherence Monitoring with 2.5D RoT}
\label{sec:CC}

While TRANSMONs are monitoring memory
transactions, they are agnostic to CC mechanisms. This is a
critical limitation as modern multicore chiplet systems are cache-coherent.
Without addressing coherence, the system ultimately
remains vulnerable to protocol-level exploits that can bypass
transaction-level defenses. To address this, our prior work on secure 2.5D
RoTs proposes embedding Coherence Message Checkers (CMCs)
directly within the physical ingress links of the active interposer's
NoC~\cite{chacon2024}.

\subsection{Coherence Message Checkers}

A CMC inspects messages traveling the NoC for
signs of malicious modification~\cite{chacon2024}. Each CMC is composed of a Packet
Checker/Modifier (PCM) and an APU (similar to that proposed
in Section~\ref{sec:APU}).
The memory regions that each chiplet is allowed to access are defined in the APU, which is managed by a secure OS running on the trusted interposer.
The PCM is responsible for validating the requested memory address against the chiplet's permissions.
If a chiplet generates a coherence
request for memory region it does not have permission to access, the request
is flagged as malicious and system execution stops.

CMCs are
implemented in two versions~\cite{chacon2024}:
\begin{itemize}
\item \textbf{CMC-1:} This CMC is located at the boundary between the
  untrusted chiplets and the interposer's NoC; it monitors and
  verifies coherence messages entering the interposer from the
  chiplets, preventing malicious coherence packets from reaching
  the network. The PCM within CMC-1 validates the requested memory
  addresses against the APU's programmed permissions. CMC-1 requires
  one pipeline stage to analyze fields in the flit and a subsequent
  pipeline stage to lookup permissions in the APU.
    
\item \textbf{CMC-2:} This CMC is physically connected to the memory
  controllers; it prevents passive snooping attacks on broadcast
  traffic.
  In broadcast-heavy protocols
  like MOESI Hammer, CMC-2's active filtering provides a unique
  defense as follows. When the memory directory broadcasts a coherence request to
  all chiplets, CMC-2 checks the memory permissions for each
  destination chiplet. If a destination 
  lacks access rights to the memory region, CMC-2 dynamically
  converts the broadcast message into a targeted
  negative-acknowledgment (NACK) or unicast response, hiding the
  transaction from unauthorized chiplets.
  CMC-2 requires an additional pipeline
  stage to filter out broadcast requests.
\end{itemize}

\subsection{Performance Overhead and Scalability}

CMC-1 increases
the queuing latency for packets entering the interposer. However, the
impact on overall system performance is modest:
for an industry-representative 64-core RISC-V system,
across single-core SPEC 2006 benchmarks, the performance
loss is approximately 4\% on average~\cite{chacon2024}. The 
overhead is strongly correlated with an application's cache
behavior: benchmarks with a low number of L2 misses 
experience minimal overhead while workloads that generate heavy
coherence traffic are more susceptible to slowdown~\cite{chacon2024}.

CMC-2's mechanism partially offsets its own latency cost, by
filtering out unnecessary broadcast messages that lack permissions for
the relevant memory region, thereby reducing the total NoC traffic. In
multi-programmed workloads, this benefit becomes more apparent with some
modes achieving higher throughput with the CMCs enabled, where
the bandwidth saved by filtering dominates the packet checking latency~\cite{chacon2024}.

The CMC design scales well when the number of nodes increases, mainly because
the CMC-1's overhead at the NoC perimeter remains constant
regardless of system size.

\section{Physical Design for 2.5D RoT}
\label{sec:EDA}

Migrating security features from monolithic SoCs to active interposers
fundamentally alters the Electronic Design Automation (EDA)
needs~\cite{park2020}.
Traditional passive interposers are analogous to
packaging substrates, designed with package-centric tools.
Active interposers, however, contain 
standard cells and require
tailored layout synthesis flows~\cite{park2020}.

In a modern 2.5D design flow proposed in our prior work, chiplets and the active
interposer are co-analyzed and co-designed~\cite{park2020}.
The flow begins with independent physical implementation of commodity chiplets in advanced nodes (e.g., 28nm).
Next, the interposer RTL is synthesized using for mature node (e.g., 65nm).
During floorplanning, the microbump
arrays of the chiplets are projected onto the interposer as I/O pins,
and C4-bumps are assigned with placement blockages for TSV pads
connecting the interposer to the package.

Aside from integrating security features, active interposers improve both Power Integrity (PI) and Signal Integrity (SI).
In a passive 2.5D design, PI may require
multiple Integrated Voltage Regulator (IVR) chiplets, which can still experience 
long power delivery paths. By embedding IVRs directly into the
active interposer, the distance between power supply and
power-demanding logic can be minimized.
In the same 64-core RISC-V system utilized before, introducing an active interposer with embedded IVRs reduced the maximum IR drop by 73.7\% compared to a passive baseline~\cite{park2020}.
Concerning SI, the insertion of repeaters into an active interposer eliminates the long, highly resistive unbuffered wires required in passive 2.5D designs, significantly mitigating signal distortion.
Furthermore, chip-centric SI tools can extract the
Standard Parasitic Exchange Format (SPEF) files from both the chiplets
and the active interposer, merging them into a single wrapper netlist
for comprehensive, system-wide SI analysis~\cite{park2020}.

Although the addition of TRANSMON/CMC security features and related modifications of the NoC routers
increases the standard-cell area within the interposer, an active 2.5D
design can actually reduce the total system silicon footprint by
18.5\% because the NoC and IVRs are removed from the surface-level
chiplets~\cite{park2020}.
The active interposer maintains a low
utilization rate of 2.68\%, making it a ``minimally active'' interposer that avoids excessive yield loss~\cite{park2020}.
The secured active design reduces total
system power by 3.2\% compared to an unsecured passive baseline,
proving that active interposers offer scalable, secure-by-construction
hardware without favorable PPA trade-offs~\cite{park2020}.

\section{Conclusion}
\label{sec:conc}

Securing modern chiplet systems demands a fundamental departure from
chip-level security paradigms. In a multi-vendor 2.5D ecosystem, any
component may originate from an untrusted
source, rendering traditional hardware security features and
RoT-as-chiplet proposals insufficient. The threat landscape spans
mainly three categories: direct interconnect attacks, CC exploits such as
complex forging attacks, and microarchitectural side-channel threats.

At design time, 2.5D split manufacturing
utilizes the interposer to hide system-level interconnects and, with
an active interposer, to partition logic across trusted and untrusted
facilities. At runtime, an active interposer serving as a physically
isolated RoT can embed TRANSMONS to enforce
memory-access permissions and CMCs to police
cache coherence, thwarting the full spectrum of identified
threats with around 4\% performance overhead and
without modifying the commodity chiplets~\cite{nabeel2020,chacon2024}.
Modern 2.5D EDA flows show that these security features can be integrated alongside
NoC routers and IVRs with net reductions in IR drop, signal
distortion, and total system silicon area~\cite{park2020}. Together,
these layered defenses provide a viable path to secure-by-construction
systems built from untrusted, multi-vendor chiplets.

In future work, additional measures such as secure boot can be integrated to further strengthen the overall security posture.
The 2.5D RoT architecture should also be tailored to support emerging interface standards like UCIe.

\section*{Acknowledgments}
\label{sec:acks}

This work was supported in part by the NYUAD Center for Cybersecurity (CCS). The authors also acknowledge the support from the Purdue Center for Secure Microelectronics Ecosystem [CSME\#210205].
\newpage
\bibliographystyle{IEEEtran}
\bibliography{references}

\end{document}